\documentclass[%
 reprint,
 amsmath,amssymb,
 aps,
 pra,
]{revtex4-2}

\usepackage{graphicx}
\usepackage{dcolumn}
\usepackage{bm}
\usepackage[colorlinks=true, linkcolor=blue, urlcolor=blue, citecolor=blue]{hyperref}
\usepackage{physics}

\begin{document}

\title{The Sound of Decoherence}

\author{Robson Christie}
 \email{robson.christie@port.ac.uk}
\affiliation{%
 School of Mathematics and Physics,\\University of Portsmouth, United Kingdom }%

\author{James Trayford}
\email{james.trayford@port.ac.uk}
\affiliation{
    Institute of Cosmology and Gravitation,\\University of Portsmouth, United Kingdom }%

\date{\today}

\begin{abstract}
We explore an unconventional bridge between quantum mechanical density matrices and sound by mapping elements of the density matrix and their phases to auditory signals, thus introducing a framework for \emph{Open Quantum Sonification}. Employing the eigenstates of the Hamiltonian operator as a basis, each quantum state contributes a frequency proportional to its energy level. The off-diagonal terms, which encode coherence and phase relationships between energy levels, are rendered as binaural signals presented separately to the left and right ears. We illustrate this method within the context of open quantum system dynamics governed by Lindblad equations, presenting first an example of quantum Brownian motion of a particle in a thermal bath, and second, a recoherence process induced by boundary driving that results in spin-helix states. This document serves as a companion to the corresponding audio visual simulations of these models available on the YouTube channel \href{https://www.youtube.com/channel/UCEAGcl4PVamWqJ5yw9PMJ1g}{Open Quantum Sonification}\footnote{\scriptsize \url{ https://www.youtube.com/channel/UCEAGcl4PVamWqJ5yw9PMJ1g}} with the Python Codes on \href{https://github.com/rchristie95/OpenQuantumSonification}{GitHub}\footnote{\scriptsize \url{https://github.com/rchristie95/OpenQuantumSonification}}. The auditory analogy presented here provides an intuitive and experiential means of describing quantum phenomena such as tunnelling, thermalisation, decoherence, and recoherence.
\end{abstract}

\maketitle

\section{Introduction}
Over the past two decades, researchers have explored \emph{Quantum Sonification} as a way to represent quantum data and dynamics using sound, with the aim of fostering intuition, educational value, or even artistic interpretation. Early efforts such as those by Sturm \cite{sturm2000sonification,sturm2001composing}, de Campo \cite{de2005sonification,vogt2007introduction,de2005sonification} and others focused on mapping quantum energy spectra directly to auditory parameters, while subsequent work expanded to include sonifications of controlled quantum dynamics \cite{kontogeorgakopoulos2014sonification,miranda2022quantum,itaborai2024developing}, immersive virtual reality environments \cite{morawitz2018quantum}, and more recently sonifications of quantum phase-space representations and Rabi oscillations \cite{yamada2023towards,yamada2024sonification}. Together, these studies illustrate a progression from static spectral mappings to dynamically evolving, context-rich auditory displays that complement traditional visualisation techniques and can reveal features like coherence, tunnelling, and non-classicality in ways that are sometimes challenging to grasp visually.

Our approach is closely aligned with the recent work of Arasaki and Takatsuka \cite{arasaki2024sonification}, who also map quantum energy levels directly onto audible frequencies. We extend this methodology to include the general framework of density matrices \cite{blum2012density}. By introducing a binaural element, assigning the \emph{ket} eigenbasis to one ear and the \emph{bra} eigenbasis to the other, we audibly emphasise the process of decoherence. In doing so, we provide a new sonic perspective on the quantum-to-classical crossover.

\section{Binaural mapping}
In quantum mechanics, the state of a system can be described by a density matrix $\hat{\rho}$, which encapsulates both the populations of energy levels and the coherence between them. The density matrix is a Hermitian operator and can be written in a chosen basis as
\begin{equation} \label{eq:rho}
    \hat{\rho}=\sum_{k,l} \rho_{k l} \ket{k}\bra{l}=\sum_{k,l} r_{k l} e^{i \theta_{k l}}\ket{k}\bra{l},
\end{equation}
where $r_{kl} \geq 0$ and $\theta_{kl}$ is the phase angle associated with the off-diagonal element $\rho_{kl}$. Hermiticity of $\hat{\rho}$ ensures
\begin{equation}
    \theta_{k l}=-\theta_{l k}.
\end{equation}
To construct a meaningful sonic analogy, we employ an \emph{additive synthesis} approach~\cite{maher1991sinewave}; we work in the basis of Hamiltonian eigenstates
\begin{equation}
    \hat H \ket{n}=E_n\ket{n},
\end{equation}
and assign a fundamental frequency $f_0$ to a pure sinusoidal tone, representing the ground state. Excited states are then mapped to overtones of $f_0$, with frequencies proportional to their energies, as follows
\begin{equation}
    f_n = \frac{E_n}{E_0} \times f_{0}
\end{equation}

The off-diagonal terms in $\hat{\rho}$ represent quantum interference between different energy levels. To capture this, we create a stereo signal in which the \emph{ket} index is directed to the left ear and the \emph{bra} index to the right ear. Thus,
\begin{equation}
    r_{k l} e^{i \theta_{k l}}\ket{k}\bra{l}
    \to \begin{cases}
    r_{k l}\sin(2 \pi f_k t+\theta_{k l}) &\text{(left ear)}, \\[6pt]
    r_{k l}\sin(2 \pi f_l t-\theta_{k l}) &\text{(right ear)}.
    \end{cases}
\end{equation}
Summing all contributions with $k\geq l$ produces a complex \emph{soundscape} that directly encodes the quantum structure and dynamics. We consider only the lower triangle of the density matrix since, due to Hermiticity, the upper triangle contains the same information with a sign-flipped phase. As these phases evolve and off-diagonal terms vanish due to decoherence, the perceived sound shifts from intricate binaural patterns to a simpler monophonic tone, mirroring the quantum-to-classical transition.

\section{\texorpdfstring{Lindblad Equations \\ and Decoherence}{Lindblad Equations and Decoherence}} \label{sec:quantum}

Realistic quantum systems rarely remain isolated; instead, they interact with external environments that induce departures from ideal unitary evolution. Such open system dynamics are captured by a \emph{master equation} that incorporates both dissipative and decoherent effects. The \emph{Lindblad form} \cite{lindblad1976generators,gorini1976completely,petruccione} specifically provides a wide class of these equations, which ensures complete positivity and trace preservation of the density matrix.

The Lindblad equation reads:
\begin{multline} \label{eq:lindblad}
\frac{d}{dt}\hat{\rho}=-\frac{i}{\hbar}[\hat{H}_\gamma,\hat{\rho}]\\ + \frac{1}{\hbar} \sum_{k}\left(\hat{L}_{k} \hat{\rho} \hat{L}_{k}^{\dagger} - \frac{1}{2}\hat{L}_{k}^{\dagger}\hat{L}_{k}\hat{\rho} - \frac{1}{2}\hat{\rho}\hat{L}_{k}^{\dagger}\hat{L}_{k}\right).
\end{multline}
Here, $\hat H_{\gamma}$ is the effective Hamiltonian and $\hat{L}_k$ are the Lindblad operators or \emph{jump operators} that encode the specific ways the environment interacts with the system.

Beyond the master equation formulation, Lindblad dynamics can be represented via \emph{Stochastic Schrödinger Equations} (SSEs) \cite{percival,belavkin1989nondemolition}, which unravel the Lindblad equation into individual quantum trajectories:
\begin{multline}\label{eq:SSE}
\ket{d \psi } = -\frac{i}{\hbar}\hat{H}_{\gamma} \ket{\psi}+\\\frac{1}{\hbar}\sum_{k}\left(-\frac{1}{2}\hat{L}_k^{\dagger}\hat{L}_k + \ev*{\hat{L}_k^{\dagger}}\hat{L}_k - \frac{1}{2}\ev*{\hat{L}_k^{\dagger}}\ev*{\hat{L}_k}\right)\ket{\psi} dt 
\\+ \frac{1}{\sqrt{\hbar}}\sum_{k}\left(\hat{L}_k-\ev*{\hat{L}_k}\right)\ket{\psi} dW_k.
\end{multline} 
where $W_k$ are independent Itô processes \cite{oksendal2013stochastic}. Averaging many such stochastic trajectories recovers the solution of the Lindblad equation:
\begin{equation}
    \hat \rho_{Lind}(t)=\mathbb{E}(\ket{\psi_{SSE}(t)}\bra{\psi_{SSE}(t)}).
\end{equation}

\subsection{Thermalisation in a Double Well}
\begin{figure}[h]
    \centering
    \includegraphics[width=0.4\textwidth]{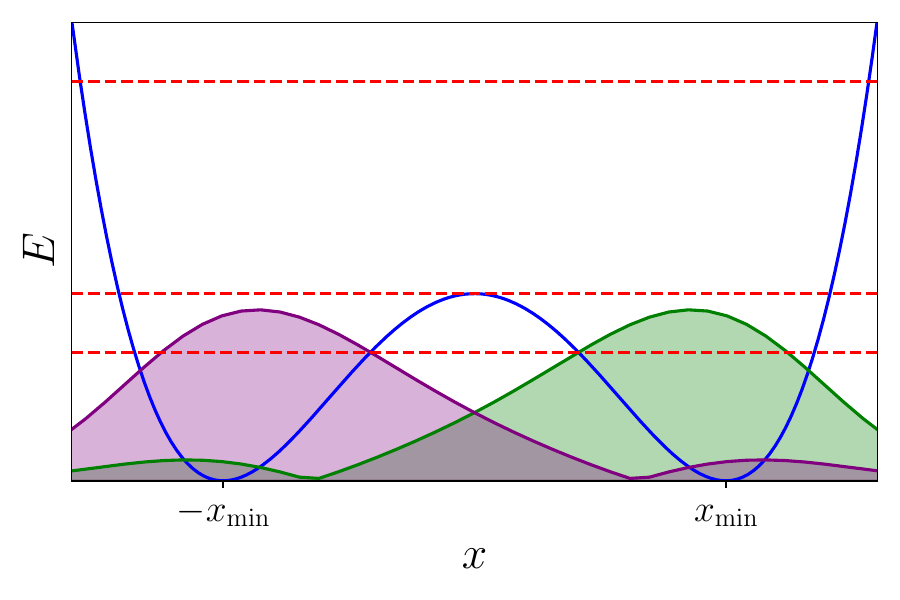}
    \caption{A double well with parameters $c_2=0.35$ and $c_2=0.05$ in blue with the first three energy levels as dashed red lines. In purple we plot the probability density of the symmetric combination of the lowest two eigenstate wave functions $\norm{\psi_0(x)+\psi_1(x)}^2/2$ in green the antisymmetric combination $\norm{\psi_0(x)-\psi_1(x)}^2/2$.}
    \label{fig:DWell}
\end{figure}
Quantum thermalisation is the process of a quantum system evolving towards thermal equilibrium through fluctuations, dissipation and decoherence. In this subsection, we investigate a thermalisation model within a double-well potential, exploring the roles of quantum tunnelling and environmental coupling.

Consider a quartic double-well potential of the form
\begin{equation} \label{eq:gausswell}
     V(\hat X)=c_4 \hat{X}^4 - c_2 \hat{X}^2,
\end{equation}
where $c_4$ and $c_2$ are constants that determine the shape and barrier height of the potential. The companion YouTube playlists showcase the \href{https://www.youtube.com/playlist?list=PLnFRudoWkGcFL1CHw-Fm1MyMWy0dQSNo1}{shallow}~\cite{shallow_double_well_playlist} and \href{https://www.youtube.com/playlist?list=PLnFRudoWkGcHVpW_9V9Xgd7ijx69aGJo8}{deep}~\cite{deep_double_well_playlist} double well simulations.  In FIG.~\ref{fig:DWell}, we illustrate a double-well potential along with its energy levels. We also demonstrate a well-known result~\cite{razavy2013quantum}: the symmetric combination of the ground and first excited states localises in one of the potential minima, while the antisymmetric combination localises in the other. At zero temperature, this double-well system supports tunnelling between its minima. Considering the time evolution of an initially symmetric combination of the ground and first excited states, we have the tunnelling dynamics
\begin{multline} \label{eq:t-tunnel} \frac{1}{\sqrt{2}} e^{-\frac{i}{\hbar}\hat{H} t}\bigl(\ket{E_0}+\ket{E_1}\bigr) = \\ \frac{1}{\sqrt{2}} e^{-\frac{i}{\hbar}E_0 t}\left(\ket{E_0} + e^{-\frac{i}{\hbar}(E_1 - E_0) t}\ket{E_1}\right), \end{multline}
with a characteristic period
\begin{equation} \label{eq:t-tunnel-period} t_{\text{tunnel}} = \frac{2\pi\hbar}{E_1 - E_0}. \end{equation}
In the sonified signal, if $E_1$ and $E_0$ are close, then this can be  perceived as a binaural `beating' (i.e. a rhythmic pulsing of the perceived composite sound) \cite{oster1973auditory} at the frequency
\begin{equation}
    f_{b}=\frac{f_0(E_1-E_0)}{2}
\end{equation}
As the barrier height increases, the tunnelling rate decreases, and consequently the perceptual beating effect decreases in frequency.

We introduce finite-temperature and coupling to the environment via a commonly used Lindblad representation derived by Petruccione in~\cite{petruccione}. This model consists of the effective Hamiltonian
\begin{equation} \label{eq:heatbathH}
    \hat{H}_{\gamma} = \frac{\hat{P}^2}{2m} + V(\hat{X}) + \frac{\gamma}{2}(\hat{X}\hat{P} + \hat{P}\hat{X}),
\end{equation}
alongside a single Lindblad jump operator
\begin{equation} \label{eq:heatbathL}
    \hat{L} = \sqrt{\frac{4 \gamma m k_B T}{\hbar}}\,\hat{X} + i\sqrt{\frac{\gamma \hbar}{4 m k_B T}}\,\hat{P},
\end{equation}
where \( \gamma \) is the damping coefficient, \( m \) is the mass of the particle, and \( \hat{X} \) and \( \hat{P} \) are the position and momentum operators, respectively. Here, \( k_B \) denotes the Boltzmann constant.

\subsection{Spin Helices in the XXZ Heisenberg Chain}
\begin{figure}[h]
    \centering
    \includegraphics[width=0.4\textwidth]{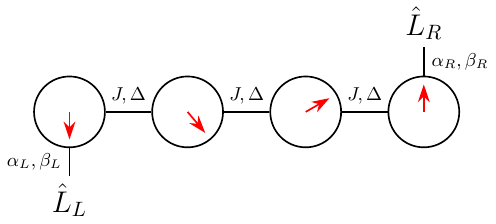}
    \caption{Illustration of a four site XXZ Heisenberg chain with boundary Lindblad operators, the red arrow represents the spin directions and the straight lines represent couplings.}
    \label{fig:4-Site}
\end{figure}
By tailoring the interactions at the boundaries of a quantum system, it is possible to counteract decoherence effects, thereby maintaining coherent quantum states essential for quantum information processing. This subsection explores how engineered boundary conditions in the XXZ Heisenberg chain can be leveraged to achieve recoherent \emph{spin helix states} \cite{kohda2012gate,kohda2017physics,jepsen2022long,kuhn2023quantum}, contrasting with the previous thermal example in which ordered tones dynamically evolved into disorder. The XXZ chain simulations are available in the \href{https://www.youtube.com/playlist?list=PLnFRudoWkGcGvAqjI4cF_ZnQsnsL0pUNZ}{YouTube playlist}~\cite{xxz_heisenberg_chain_playlist}.

As a test bed for the dynamical generation of spin helix states we use the Markovian model previously derived by Popkov and Sch\"utz in \cite{popkov2017solution}, consisting of a one-dimensional chain of spin-$\tfrac{1}{2}$ particles subject to an XXZ Heisenberg Hamiltonian
\begin{equation} \label{eq:XXZ_H}
    \hat{H} = J \sum_{j=1}^{N-1} \left( \hat{\sigma}_j^x \hat{\sigma}_{j+1}^x + \hat{\sigma}_j^y \hat{\sigma}_{j+1}^y + \Delta\bigl(\hat{\sigma}_j^z \hat{\sigma}_{j+1}^z - \hat{I}\bigr) \right),
\end{equation}
where \( J \) is the exchange coupling, \(\Delta = \cos(\eta)\) is the anisotropy parameter, and \(\hat{\sigma}_j^{x,y,z}\) are Pauli matrices. This Hamiltonian supports a range of nontrivial spin configurations, including spin helix states, which arise when the spin direction smoothly twists along the chain.

To stabilise a spin helix state as a non-equilibrium steady state, we couple the chain’s boundary spins to an environment using specifically engineered Lindblad operators \cite{popkov2017solution}. These operators enforce a phase-coherent pattern of spin raising and lowering processes at the edges:
\begin{align} \label{eq:D_L}
    \hat{L}_L &= \alpha_L(r \hat{\sigma}_1^- \hat{\sigma}_1^+) - \beta_L\left(\frac{\hat{\sigma}_1^z - \hat{I}}{2} - r \hat{\sigma}_1^-\right),\\ \label{eq:D_R}
    \hat{L}_R &= \alpha_R\bigl(r e^{-i\Phi} \hat{\sigma}_N^- \hat{\sigma}_N^+\bigr) - \beta_R\left(\frac{\hat{\sigma}_N^z - \hat{I}}{2} - r e^{i\Phi}\hat{\sigma}_N^-\right).
\end{align}
Here, \(\alpha_{L,R}\), \(\beta_{L,R}\), \(r\), and \(\Phi\) control the injection and extraction rates of the boundary spin and set the spatial twisting of the spin phase. The resulting dissipative dynamics, described by the Lindblad equation, guide the chain into a pure spin helix state. We illustrate this set-up in FIG~\ref{fig:4-Site}.

When sonified, the evolving local spin observables produce a distinct audio signal that differs markedly from uniform or random spin distributions. As the helix forms and stabilises, the auditory pattern reflects the emergence of a coherent, smoothly varying spin pattern, as a direct acoustic signature of this spatially modulated quantum state.

\section{Outlook and Applications}
Quantum sonification provides an innovative interpretive and educational framework, enabling researchers and students to perceive the transition from quantum to classical regimes through sound. By making these inherently abstract phenomena audible, it offers an immersive, intuitive complement to traditional visualization methods. Beyond simple systems, this approach can be extended to many-body states, entanglement, and even topological phases of matter. Ultimately, the ability to “hear” quantum states can inspire more nuanced interpretations, guide exploratory research in uncharted quantum territories, and enrich our understanding of the quantum world.

\section*{Acknowledgments}
We wish to thank Jaewoo Joo for the suggestion to include an example on spin helix states. JT acknowledges the support of UKRI STFC, \textit{Early Stage Research \& Development Grant}, reference ST/X004651/1.

\bibliographystyle{ieeetr}
\bibliography{bib}

\begin{thebibliography}{10}

\bibitem{sturm2000sonification}
B.~L. Sturm, ``Sonification of particle systems via de broglie's hypothesis,'' in {\em Proceedings of the International Conference on Auditory Display (ICAD)}, Georgia Institute of Technology, 2000.

\bibitem{sturm2001composing}
B.~L. Sturm, ``Composing for an ensemble of atoms: the metamorphosis of scientific experiment into music,'' {\em Organised Sound}, vol.~6, no.~2, pp.~131--145, 2001.

\bibitem{de2005sonification}
A.~de~Campo, C.~Frauenberger, R.~H{\"o}ldrich, T.~Melde, W.~Plessas, and B.~Sengl, ``Sonification of quantum spectra,'' in {\em Proceedings of ICAD}, pp.~6--9, 2005.

\bibitem{vogt2007introduction}
K.~Vogt, A.~de~Campo, and G.~Eckel, ``An introduction to sonification and its application to theoretical physics,'' in {\em Proceedings of the 3rd Congress of the Alps Adria Acoustics Association, Graz, Austria}, 2007.

\bibitem{kontogeorgakopoulos2014sonification}
A.~Kontogeorgakopoulos and D.~Burgarth, ``Sonification of controlled quantum dynamics,'' in {\em ICMC}, 2014.

\bibitem{miranda2022quantum}
E.~R. Miranda, ``Quantum computer music,'' {\em Foundations, Methods and Advanced Concepts. Springer, Cham}, 2022.

\bibitem{itaborai2024developing}
P.~V. Itabora{\'\i}, P.~Thomas, A.~Crippa, K.~Jansen, T.~Schw{\"a}gerl, and M.~A. Y{\'a}{\~n}ez, ``Developing a framework for sonifying variational quantum algorithms: Implications for music composition,'' {\em arXiv preprint arXiv:2409.07104}, 2024.

\bibitem{morawitz2018quantum}
F.~Morawitz, ``Quantum: An art-science case study on sonification and sound design in virtual reality,'' in {\em 2018 IEEE 4th VR Workshop On Sonic Interactions For Virtual Environments (SIVE)}, pp.~1--5, IEEE, 2018.

\bibitem{yamada2023towards}
R.~Yamada, E.~Pi{\~n}ol, S.~Grandi, J.~Zakrzewski, and M.~Lewenstein, ``Towards the intuitive understanding of quantum world: Sonification of {R}abi oscillations, {W}igner functions, and quantum simulators,'' {\em arXiv preprint arXiv:2311.13313}, 2023.

\bibitem{yamada2024sonification}
R.~Yamada, A.~Reserbat-Plantey, E.~Pi{\~n}ol, and M.~Lewenstein, ``Sonification of {W}igner functions: case study of intense light-matter interactions,'' in {\em International Conference on Mathematics and Computation in Music}, pp.~447--455, Springer, 2024.

\bibitem{arasaki2024sonification}
Y.~Arasaki and K.~Takatsuka, ``Sonification of molecular electronic energy density and its dynamics,'' {\em RSC advances}, vol.~14, no.~13, pp.~9099--9108, 2024.

\bibitem{blum2012density}
K.~Blum, {\em Density matrix theory and applications}, vol.~64.
\newblock Springer Science \& Business Media, 2012.

\bibitem{maher1991sinewave}
R.~C. Maher, ``Sinewave additive synthesis revisited,'' in {\em Audio Engineering Society Convention 91}, Audio Engineering Society, 1991.

\bibitem{lindblad1976generators}
G.~Lindblad, ``On the generators of quantum dynamical semigroups,'' {\em Communications in Mathematical Physics}, vol.~48, pp.~119--130, 1976.

\bibitem{gorini1976completely}
V.~Gorini, A.~Kossakowski, and E.~Sudarshan, ``Completely positive dynamical semigroups of {N}-level systems,'' {\em Journal of Mathematical Physics}, vol.~17, p.~821, 1976.

\bibitem{petruccione}
H.~Breuer and F.~Petruccione, {\em The Theory of Open Quantum Systems}.
\newblock Oxford University Press, 2002.

\bibitem{percival}
I.~Percival, {\em Quantum State Diffusion}.
\newblock Cambridge University Press, 1998.

\bibitem{belavkin1989nondemolition}
V.~Belavkin, ``A new wave equation for a continuous nondemolition measurement,'' {\em Physics Letters A}, vol.~140, no.~5-6, pp.~355--358, 1989.

\bibitem{oksendal2013stochastic}
B.~Oksendal, {\em Stochastic differential equations: an introduction with applications}.
\newblock Springer Science \& Business Media, 2013.

\bibitem{shallow_double_well_playlist}
{Open Quantum Sonification}, ``Shallow double well playlist.'' \url{https://www.youtube.com/playlist?list=PLnFRudoWkGcFL1CHw-Fm1MyMWy0dQSNo1}, 2024.

\bibitem{deep_double_well_playlist}
{Open Quantum Sonification}, ``Deep double well playlist.'' \url{https://www.youtube.com/playlist?list=PLnFRudoWkGcHVpW_9V9Xgd7ijx69aGJo8}, 2024.

\bibitem{razavy2013quantum}
M.~Razavy, {\em Quantum theory of tunneling}.
\newblock World Scientific, 2013.

\bibitem{oster1973auditory}
G.~Oster, ``Auditory beats in the brain,'' {\em Scientific American}, vol.~229, no.~4, pp.~94--103, 1973.

\bibitem{kohda2012gate}
M.~Kohda, V.~Lechner, Y.~Kunihashi, T.~Dollinger, P.~Olbrich, C.~Sch{\"o}nhuber, I.~Caspers, V.~Bel'Kov, L.~Golub, D.~Weiss, {\em et~al.}, ``Gate-controlled persistent spin helix state in (in, ga) as quantum wells,'' {\em Physical Review B—Condensed Matter and Materials Physics}, vol.~86, no.~8, p.~081306, 2012.

\bibitem{kohda2017physics}
M.~Kohda and G.~Salis, ``Physics and application of persistent spin helix state in semiconductor heterostructures,'' {\em Semiconductor Science and Technology}, vol.~32, no.~7, p.~073002, 2017.

\bibitem{jepsen2022long}
P.~N. Jepsen, Y.~K. Lee, H.~Lin, I.~Dimitrova, Y.~Margalit, W.~W. Ho, and W.~Ketterle, ``Long-lived phantom helix states in {H}eisenberg quantum magnets,'' {\em Nature Physics}, vol.~18, no.~8, pp.~899--904, 2022.

\bibitem{kuhn2023quantum}
S.~K{\"u}hn, F.~Gerken, L.~Funcke, T.~Hartung, P.~Stornati, K.~Jansen, and T.~Posske, ``Quantum spin helices more stable than the ground state: Onset of helical protection,'' {\em Physical Review B}, vol.~107, no.~21, p.~214422, 2023.

\bibitem{xxz_heisenberg_chain_playlist}
{Open Quantum Sonification}, ``{XXZ} {H}eisenberg chain playlist.'' \url{https://www.youtube.com/playlist?list=PLnFRudoWkGcGvAqjI4cF_ZnQsnsL0pUNZ}, 2024.

\bibitem{popkov2017solution}
V.~Popkov and G.~M. Sch{\"u}tz, ``Solution of the {L}indblad equation for spin helix states,'' {\em Physical Review E}, vol.~95, no.~4, p.~042128, 2017.

\end{thebibliography}

\end{document}